\documentclass[12pt]{iopart}

\usepackage{iopams}

\usepackage{graphicx}
\usepackage{txfonts,color}
\usepackage{ulem}
\usepackage{booktabs}
\usepackage{dsfont}

\newcommand{\ket}[1]{| #1 \rangle}

\usepackage[unicode=true,pdfusetitle,
 bookmarks=false,
 breaklinks=false,pdfborder={0 0 1},backref=false,colorlinks=false]
 {hyperref}
\usepackage[dvipsnames]{xcolor}
 \hypersetup{colorlinks=true,citecolor=NavyBlue,filecolor=black,linkcolor=black,urlcolor=black}
 
\makeatother

\usepackage{graphicx}
\usepackage{ulem}
\usepackage{txfonts,color}
\usepackage{dsfont}
\usepackage{booktabs} 
\makeatother

\usepackage{iopams}

\begin{document}

\title[Neural networks for Bayesian quantum many-body magnetometry]{Neural networks for Bayesian quantum many-body magnetometry}

\author{Yue Ban$^{1}$, Jorge Casanova$^{2,3,4}$ and Ricardo Puebla$^{5}$ }
\address{$^1$ TECNALIA, Basque Research and Technology Alliance (BRTA), 48160 Derio, Spain}
\address{$^2$ Department of Physical Chemistry, University of the Basque Country UPV/EHU, Apartado 644, 48080 Bilbao, Spain}
\address{$^3$ EHU Quantum Center, University of the Basque Country UPV/EHU, Leioa, Spain}
\address{$^4$ IKERBASQUE,  Basque  Foundation  for  Science, Plaza Euskadi 5, 48009 Bilbao,  Spain}
\address{$^5$ Departamento de F{\'i}sica, Universidad Carlos III de Madrid, Avda. de la Universidad 30, 28911 Legan{\'e}s, Spain}

\begin{abstract}
Entangled quantum many-body systems can be used as sensors that enable the estimation of parameters with a precision larger than that achievable with ensembles of individual quantum detectors. Typically, the parameter estimation strategy requires the microscopic modelling of  the quantum many-body system, as well as a an accurate description of its dynamics. This entails a complexity that can hinder the applicability of Bayesian inference techniques. In this work we show how to circumvent these issues by using neural networks that faithfully reproduce the dynamics of quantum many-body sensors, thus allowing for an efficient Bayesian analysis. We exemplify with an XXZ model driven by magnetic fields, and show that our method is capable to yield an estimation of field parameters beyond the standard quantum limit scaling. Our work paves the way for the practical use of quantum many-body systems as black-box sensors exploiting quantum resources to improve precision estimation.
\end{abstract}
	
\maketitle

\section{Introduction}
The key objective of quantum metrology~\cite{Giovannetti04,Giovannetti06,Giovannetti11} consists in  enhancing the precision of measured physical parameters exploiting quantum resources. In this context, the quantum Fisher information plays a central role as it sets a bound on the maximum achievable precision of the target parameter~\cite{Braunstein94,Paris09}. 
In particular, it has been shown that a quantum entangled many-body system acting as a quantum sensor offers advantages with respect to individual quantum registers: An entangled state of $N$ quantum particles can reach the so-called Heisenberg limit where the precision scales as $1/N$~\cite{Giovannetti11,Giovannetti04,Giovannetti06,Leibfried04} rather than $1/\sqrt{N}$ that corresponds to the standard quantum limit. Note, however, that decoherence effects can push the precision scaling back to the standard quantum limit~\cite{Huelga97}. Remarkably, for specific scenarios and interactions  even better precision scalings have been identified~\cite{Boixo07,Roy08,Pang17}.  Hence, devising optimal quantum metrological protocols to boost parameter precision is inherently linked with quantum many-body physics~\cite{Giovannetti11,Jones20,Montenegro22,Mishra22,Yang22}.  This is particularly relevant for critical systems, i.e. systems featuring a continuous phase transition, where the estimation benefits from the high sensitivity of the system close to its critical point~\cite{Zanardi08,Boyajian16,FernandezLorenzo17,Rams18,Garbe20,Ivanov20,Chu21,Gietka22,Garbe22,Garbe22b,Ilias22}.


In another vein, machine learning (ML)~\cite{Bishop} has become an effective and increasingly popular tool to address complex problems in distinct fields of modern quantum technologies~\cite{Carleo19,Torlai20rev,Krenn22}. Assisted by ML, several protocols have been put forward for quantum phase estimation~\cite{Lumino18,Xiao19,Palittapongarnpim19}, parameter estimation~\cite{Xu19,Peng20,Schuff20,Fiderer21,Xiao22,Madsen21}, calibration of quantum sensors~\cite{Cimini19,Nolan21}, quantum optimal control~\cite{Niu19,Bukov18,Brown21,Borah21,Porotti22,Ding21,Ai22}, and even to design quantum experiments~\cite{Krenn16,Melnikov18,Krenn20,Cervera22}, among many others. For example, in the context of quantum sensing, neural networks (NNs) allow for the use of quantum sensors even in regimes where the output response becomes complex and even in the presence of large shot noise, thus extending its working regime~\cite{Ban21,Chen22}. Employing a NN that replaces the computation of a physical observable offers a drastic reduction in the amount of sample statistic~\cite{Torlai20}.

On the other hand, Bayesian inference schemes~\cite{Linden,Gelman} enable the accurate estimation of external parameters from a complex sensor response in quantum systems~\cite{Puebla21}. The combination of ML techniques for Bayesian analysis  have led to powerful inference protocols~\cite{Fiderer21}. However, in the context of many-body system, Bayesian inference poses a challenge as only very specific many-body systems admit a closed form solution. This is so because the computation  of the dynamics of a quantum many-body system can be a daunting task even for few spins, and totally unfeasible for larger systems.

In this paper, we design a quantum many-body magnetometer assisted by NNs that enable Bayesian analysis. We illustrate our protocol for a XXZ model that acts as a magnetometer, and show that target magnetic fields can be detected efficiently with high precision (i.e. beyond the standard quantum limit scaling) via Bayesian inference. Playing the role in constructing the likelihood, NNs reproduce the microscopic dynamics with high accuracy, even with the consideration of shot-noise measurements and decoherence. Importantly, the training of the NNs could be performed solely based on experimental data, thus opening the door to employ experimental quantum many-body systems as sensors that are beyond the classical simulation capabilities.  Our results demonstrate that the super-Heisenberg scaling reported in~\cite{Rams18} is recovered when replacing the exact dynamics by the output provided by a suitably trained NN. In Sec.~\ref{s:Bayes}, we introduce our protocol of Bayesian inference which utilizes NNs to do parameter estimation. In Sec.~\ref{s:NN}, we elaborate the approximated dynamics with the application of NNs for magnetic fields with one or two vector components. In Sec.~\ref{s:scaling}, we analyze the precision scaling law with the system size. Finally, we present the conclusions in Sec.~\ref{s:con}.

\section{Bayesian inference assisted by neural networks}\label{s:Bayes}
In this Section, we introduce Bayesian inference of parameters assisted by a NN. Bayesian inference is a method of statistical inference in which Bayes' theorem is used to update the probability distributions over a set of $M$ target parameters $\Theta={\theta_1,\theta_2,\ldots, \theta_M}$ that we aim to estimate from a certain prior hypothesis $P(\Theta)$ given a set of experimental observations $\textbf{D}$. That is, one can apply Bayes’ theorem to find the posterior distribution 
\begin{eqnarray}
\label{Bayes}
P(\Theta | \textbf{D}) \propto P(\textbf{D} | \Theta)  P(\Theta),
\end{eqnarray}
where $P(\textbf{D} | \Theta)$ denotes the likelihood that such observations $\textbf{D}$ correspond to a set of parameters $\Theta$. Let us stress that $\textbf{D}$ denotes the experimental data from which we aim to obtain the information of the parameters $\Theta$. For a particular case in which the observations $\textbf{D}$ are the result of measurements on a qubit, the outcomes can be either $0$ or $1$ for the qubit being in a suitably chosen basis $\ket{0}$ or $\ket{1}$, i.e. $\textbf{D}=\left\{ X_j\right\}$ with $j=1,\ldots, N_T$ where $X_j$ denotes the number of successes obtained after measuring $N_m$ times at time $t_j$. That is, the values $X_j$ from $N_m$ measurements at each time $t_j$ form the data $\textbf{D}$. In this manner, the likelihood can be written as 
\begin{eqnarray}
\label{likelihood-original}
P(\textbf{D} | \Theta) = \Pi_{j=1}^{N_T} f(X_j, N_m, p(\Theta)).
\end{eqnarray}
The function $f$ denotes the probability of having observed exactly $X_j$ success outcomes from $N_m$ trials with success probability $p(\Theta)$, i.e.  Bernoulli distribution, where $f(X_j, N_m, p) = N_m! / (X_j!(N_m-X_j)!) p^{X_j} (1-p)^{N_m-X_j}\propto p^{X_j}(1-p)^{N_m-X_j}$. Note that the success probability depends on the target parameters $\Theta$, which in turn depends on the specific microscopic Hamiltonian model $\hat{H}(\Theta)$ that governs the dynamics of the system. Without lack of generality, let us assume that we monitor an observable $\hat{A}$, which corresponds to the probability of finding the state of a qubit in the upper state $\ket{1}$~\footnote{Trivially, one could consider other observables by taking into account  its  corresponding probability $p(\Theta)$ in the likelihood.}
\begin{eqnarray}
\label{A}
\langle \hat{A}(t ; \Theta) \rangle = {\rm{Tr}}  \left[\hat{A} \hat{U}(t) \hat{\rho}_0 \hat{U}^\dagger(t)  \right],
\end{eqnarray}
where $\hat{U}(t) = e^{-i \hat{H}(\Theta)t}$ is the time evolution operator of the whole system under the target parameters $\Theta$ and $\hat{\rho}_0$ denotes the initial state ($\hbar=1$). In this manner, the likelihood can be written as
\begin{eqnarray}
\label{PDTheta}
P(\textbf{D} | \Theta) = \Pi_{j=1}^{N_T} f(X_j, N_m, \langle A (t_j; \Theta) \rangle).
\end{eqnarray}
from where the posterior follows (cf. Eq. (\ref{Bayes})), and thus the estimation of the target parameters.

At this stage it is important to stress that the computation of $\langle \hat{A}(t;\Theta)\rangle$ poses a challenge to perform Bayesian inference on generic quantum many-body systems. Indeed, Bayesian inference requires the  computation of  $\langle \hat{A}(t;\Theta)\rangle$ for each of the candidate values $\Theta$. Since integrable quantum many-body systems are scarce, a closed-form solution for $\langle \hat{A}(t;\Theta)\rangle$ is not available. Thus, obtaining Eq.~(\ref{PDTheta}) leads to a large bottleneck, if not completely unfeasible even for moderate system sizes, which hinders the application of estimation protocols for general quantum many-body systems. Moreover, note that $\langle \hat{A}(t;\Theta)\rangle$ requires a microscopic modelling $\hat{H}(\Theta)$ that accurately describes the system including all sort of potential experimental imperfections, as otherwise the target parameters may be poorly inferred.

\begin{figure}[h]
	\begin{center}
		\scalebox{0.25}[0.25]{\includegraphics{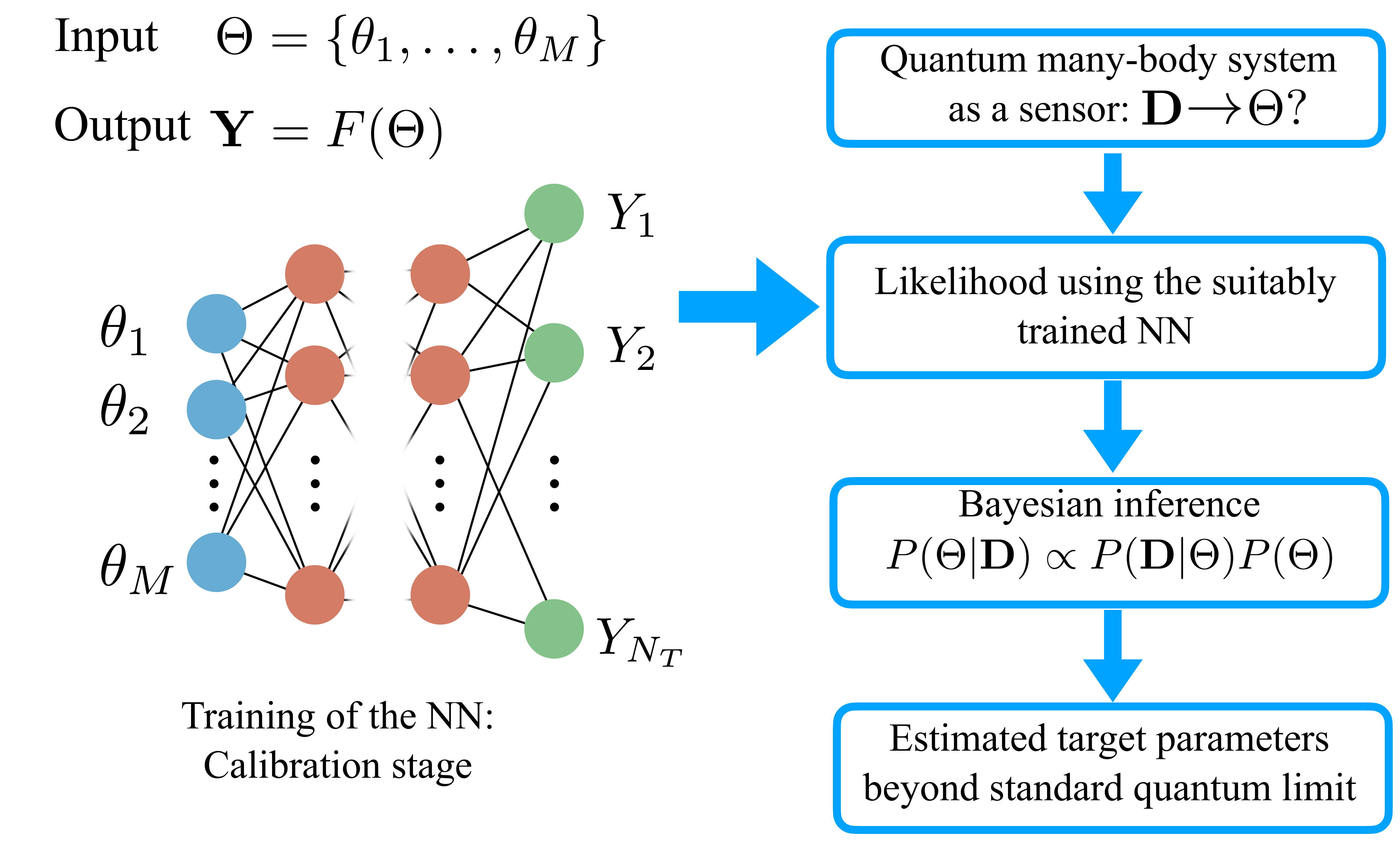}}
		\caption{\label{Fig1}  Sketch of the estimation strategy for quantum many-body systems. The calibration stage consists in training a NN that faithfully reproduces the dynamics of a chosen observable $\hat{A}$ for different candidate target parameters $\Theta$. This allows to use  the quantum many-body system as a block-box sensor, lifting the need for an accurate microscopic model and the computation of its dynamics, which becomes unfeasible even for a few number of particles. Upon calibration, a set of observations on the quantum many-body system ${\bf D}$ together with the  NN enables an efficient computation of the likelihood, and so the estimation of target parameters via Bayesian inference. This hybrid scheme can improve the precision with respect to the standard quantum limit. }
	\end{center}
\end{figure}

Fortunately, NNs can  help to simplify the computation and reproduce the microscopic dynamics with high accuracy. Let us denote $F(\Theta)$ the action of a NN which takes a parameter array $\Theta$ of dimension $M$ as input, so that $F(\Theta)$ outputs the an array ${\bf Y}$ of dimension $N_T$ corresponding to the expectation value of the chosen observable $\hat{A}$ at times $t_j$ ($j=1, ..., N_T$). The suitably trained NN establishes the relation  between the inputs $\Theta = \{\theta_1, ..., \theta_M\}$ and the outputs ${\bf Y}$, i.e. we aim at finding
\begin{eqnarray}
\label{F-A}
{\bf Y}_{j=1,\ldots,N_T}\equiv F(\Theta)_{j=1,\ldots, N_T} \approx \langle \hat{A}(t_{j=1, \ldots, N_T}; \Theta)\rangle 
\end{eqnarray}
within a tolerant error. 
As a consequence, the likelihood can be rewritten as 
\begin{eqnarray}
\label{likelihood-NN}
P(\textbf{D} | \Theta) = \Pi_{j=1}^{N_T} f(X_j, N_m, F(\Theta)_{j, \ldots, N_T}).
\end{eqnarray}
Once $F(\Theta)$ is established by the NN, its evaluation is highly efficient, which significantly decreases the computational cost. As customary in hybrid quantum-classical algorithms, the observations ${\bf D}$ stem from measurements on a quantum system, while the data processing is done in classical computer. In Fig.~\ref{Fig1}, we show a sketch of our hybrid estimation scheme, where NNs play the role of imitating the microscopic dynamics, that enables a fast computation of the likelihood and thus to do parameter estimation. Note that the NN is able to reproduce the microscopic dynamics of the many-body system for a chosen observable, taken into account all experimental imperfections and for any system size. Lifting the requirement of possessing an accurate microscopic description $\hat{H}(\Theta)$, this hybrid scheme comes at the cost of a calibration step, where the many-body system response to different candidate target parameters $\Theta$ must be scanned to then train the NN. Such calibration stage may be repeated at later times to ensure a proper functioning of the NN. Upon the calibration, the considered quantum many-body system can be used as black-box sensor allowing for an efficient Bayesian parameter estimation, while still profiting from quantum resources to boost the precision and overcoming the unfeasible computation of the dynamics for moderate and large system sizes. Note, however, that an improvement in the precision thanks to quantum resources depends on different factors~\cite{Giovannetti11}, such as the initial state $\hat{\rho}_0$, measured observable, specific quantum many-body system $\hat{H}(\Theta)$, experimental imperfections and decoherence.

\begin{figure}[t]
	\begin{center}
		\scalebox{0.28}[0.28]{\includegraphics{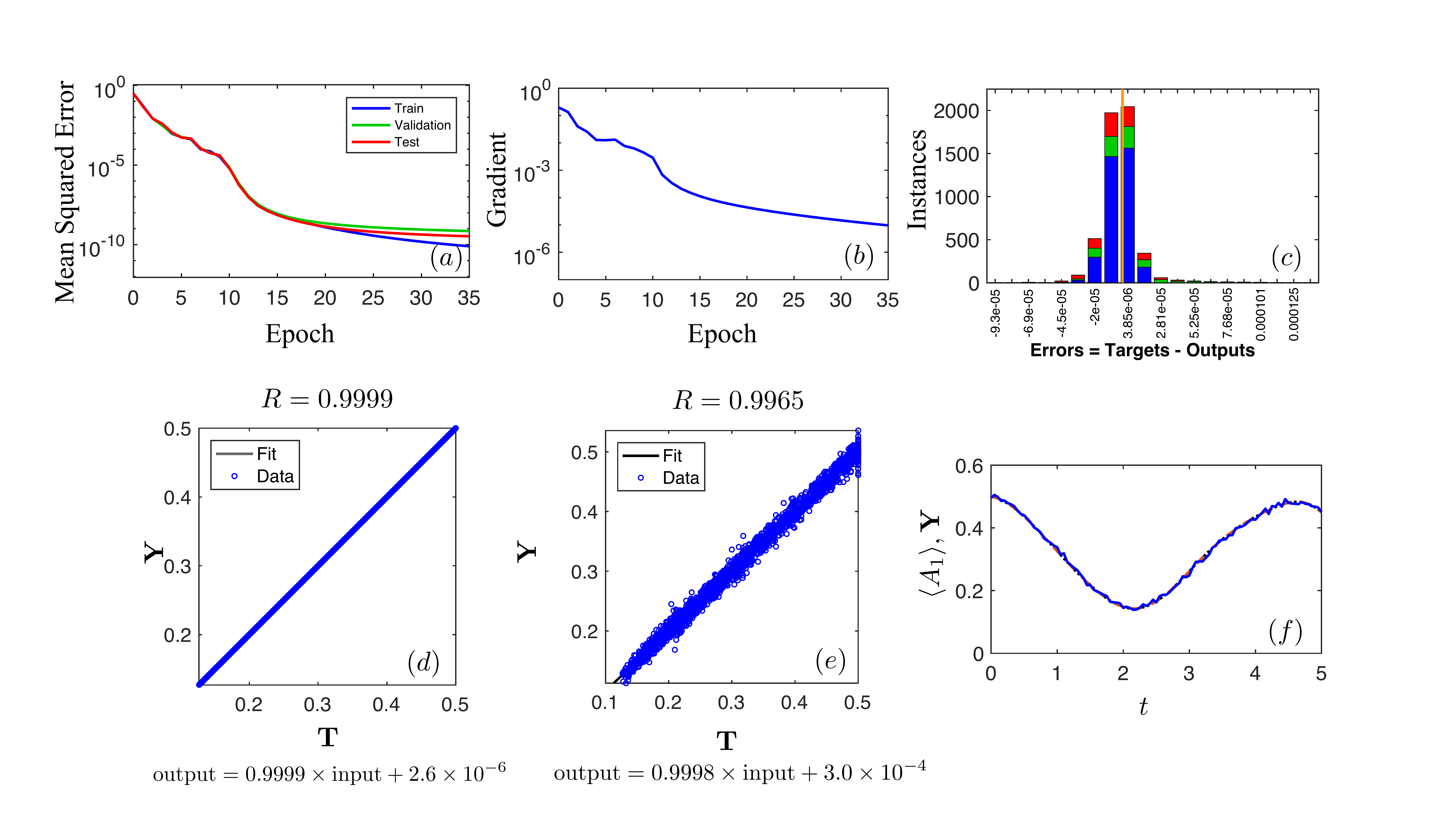}}
		\caption{\label{Fig2} (a-d) Training results of the NN to establish the relation 
$F_1(g_x)_{j=1, ..., N_T} \approx \langle \hat{A}_1(t_{j=1, ..., N_T}; g_x)\rangle$ in the presence of 1D magnetic field using a XXZ model with $N=8$ spins. The NN is trained based on the data directly derived from the exact expectation value of the observable $\hat{A}_1$: (a) Cost function or mean squared error; (b) gradient of the training set at different epoches; at the 35th epoch (c) histogram and (d) Regression of the outputs of the NN trained by the data derived without shot noise with respect to the ideal targets $\textbf{T} = \{\langle \hat{A}_1(t_{1}; g_x)\rangle, ..., \langle \hat{A}_1(t_{101}; g_x)\rangle \}$. (e)  Regression of the outputs of the NN trained based on the data obtained from the numerical simulation including finite number of measurements, here $N_m=100$, with respect to the ideal targets $\textbf{T}$.
In both (d) and (e), the fit lines (solid black) given in the equations shown below each plot, almost overlap with the line $\textbf{Y} = \textbf{T}$, while $R$ is the correlation coefficient of the outputs and the targets. (f) Comparison of the outputs from the NNs trained by the data derived without shot noise (dashed red) and including shot noise $N_m=100$ (solid blue) in accordance with the numerical ideal results $\langle \hat{A}_1(t_{j=1, ..., N_T}; g_x) \rangle$ from the Hamiltonian $\hat{H}_1$ (dotted black), where $g_x = 0.1253$. }
	\end{center}
\end{figure}

\section{Simulation of quantum many-body dynamics by neural networks}\label{s:NN}
In order to illustrate the capability of NNs to reproduce the dynamics of a chosen observable for a quantum many-body system, we consider a XXZ spin-1/2 chain in the presence of an external field that we aim to estimate. We stress that the specific many-body system is irrelevant for the proposed hybrid estimation protocol. However, we choose the XXZ spin-1/2 chain since this model has been shown to display a precision scaling beyond the quantum standard limit, see Ref.~\cite{Rams18}. This quantum many-body system constitutes therefore a good example to test whether the proposed estimation strategy can still render a quantum enhancement, which we will discuss in Sec.~\ref{s:scaling}. In the following we show two examples to test the performance of NNs to reproduce the quantum many-body dynamics of interest, namely, for a one and two dimensional external field. In addition, in \ref{app:decoh} we show how the NNs are able to accurately reproduce the dynamics under potential experimental imperfections, such as when decoherence effects are present. 


\subsection{1D external magnetic field} \label{1D}
We first start from the ferromagnetic XXZ spin-1/2 for $N$ particles with open boundary conditions as a quantum many-body system in the presence of a 1D external field. The Hamiltonian of the system read as
\begin{eqnarray}
\label{H-1D}
\hat{H}_1(g_x) = - \sum_{i=1}^{N-1} (\hat\sigma_i^x \hat\sigma_{i+1}^x + \hat\sigma_i^y \hat\sigma_{i+1}^y+J_z\sigma_i^z\sigma_{i+1}^z) + g_x\sum_{i=1}^N \hat\sigma_i^x.
\end{eqnarray}
This model displays a critical point at $g_x=0$~\cite{Affleck99,Rams18}.
In the following we consider the initial state $\hat{\rho}_0$ as the ground state without the external field $g_x=0$, i.e. at the critical point. Moreover, we take $J_z=0$ as considered in Ref.~\cite{Rams18}. The strength of the external field can  be estimated by measuring the response delivered by the system, in particular, we choose a local observable $ \hat{A}_1 = (\hat\sigma_{N/2}^x + 1) /2 $, so that $\Theta=\{ g_x\}$.
Our aim is to build a NN that encodes the relation 
\begin{eqnarray}
\label{F-A-gx}
F_1(g_x)_{j=1, ..., N_T} \approx \langle \hat{A}_1(t_{j=1, ..., N_T}; g_x)\rangle
\end{eqnarray}
from a finite number of measurements, i.e. without having access to the exact expectation value $\langle \hat{A}_1(t_{j=1, ..., N_T}; g_x)\rangle$ as it would be the case in any experimental realization. This corresponds to the calibration stage in Fig.~\ref{Fig1}.

Yet, before proceeding further, it is convenient to first consider an ideal scenario where our NN can be trained using the exact expectation value $\langle \hat{A}_1(t_{j=1, ..., N_T}; g_x)\rangle$. This will allow us to quantify the deviations from this ideal scenario when training the NN with shot noise measurements. To generate the dataset, we derive the expectation value $\langle \hat{A}_1(t_{j=1, ..., N_T}; g_x)\rangle$ as the target data  during the time interval $ 0 \leq t \leq t_f= 5$ at $N_T =101$ time instants, i.e., $t \in \{0, 0.05, ...,5\}$. As the inputs of the NN, the magnetic field is in the range $0 \leq g_x \leq 0.5$ with $51$ examples such that $ g_x \in \{0, 0.01, ...,0.5\}$. The whole dataset is then composed by $n=51$  examples, each of which has the output data $\textbf{Y} = \{y_1, y_2, ..., y_{101}\}$ approaching the targets  $\textbf{T} =\{a_1, ..., a_{101}\} = \{ \langle \hat{A}_1(t_1;g_x)\rangle, ...,  \langle \hat{A}_1(t_{101}; g_x)\rangle \}$ --a string of $101$ numbers-- while the input data is $\Theta = \{g_x\}$. The training/validation/test datasets are built with the data randomly chosen $70\%$ / $15\%$ / $15\%$ from the total dataset. We introduce the cost function as the mean squared error,
\begin{eqnarray}\label{C}
C = \sum_{i=1}^n  \sum_{j=1}^{N_T} \frac{1}{nN_T} (y_i^j -a_i^j)^2.
\end{eqnarray}
The NN is adjusted by minimizing the cost function for a training set with $n$ examples employing gradient descent, that ultimately leads to $F_1(g_x)_{j=1, ..., N_T} \approx \langle \hat{A}_1(t_{j=1, ..., N_T}; g_x)\rangle$. 
We take a XXZ system with $N=8$ spins as an example and measure the expectation value $\langle \hat{A}_1 \rangle =  (\langle\hat\sigma_{4}^x \rangle+ 1) /2 $.
Tuning the parameters of this NN with $1$ input neuron and $101$ output neurons, we find that four hidden layers with $6, 12, 25, 50$ neurons with a hyperbolic-tangent activation function~\cite{Bishop} give the highest accuracy. The training of such a NN stops when $C<10^{-5}$ at the 35th epoch leading to the results plotted in Fig.~\ref{Fig2} (a)-(d). The regression of the outputs of the NN with respect to the targets $\textbf{T}$ derived from numerical simulation is shown in Fig.~\ref{Fig2} (d). As an illustration to corroborate the good performance of the NN in reproducing the dynamics of the system, in Fig.~\ref{Fig2} (f), we show the exact expectation value $\langle \hat{A}_1(t_{j=1, ..., N_T}; g_x)\rangle$ (dotted black) for a randomly chosen example $g_x = 0.1253$ --that does not belong to the training/validation/test datasets-- and the output of the NN (dashed red), which overlap to a very good approximation. 

\begin{figure}[t]
	\begin{center}
		\scalebox{0.3}[0.3]{\includegraphics{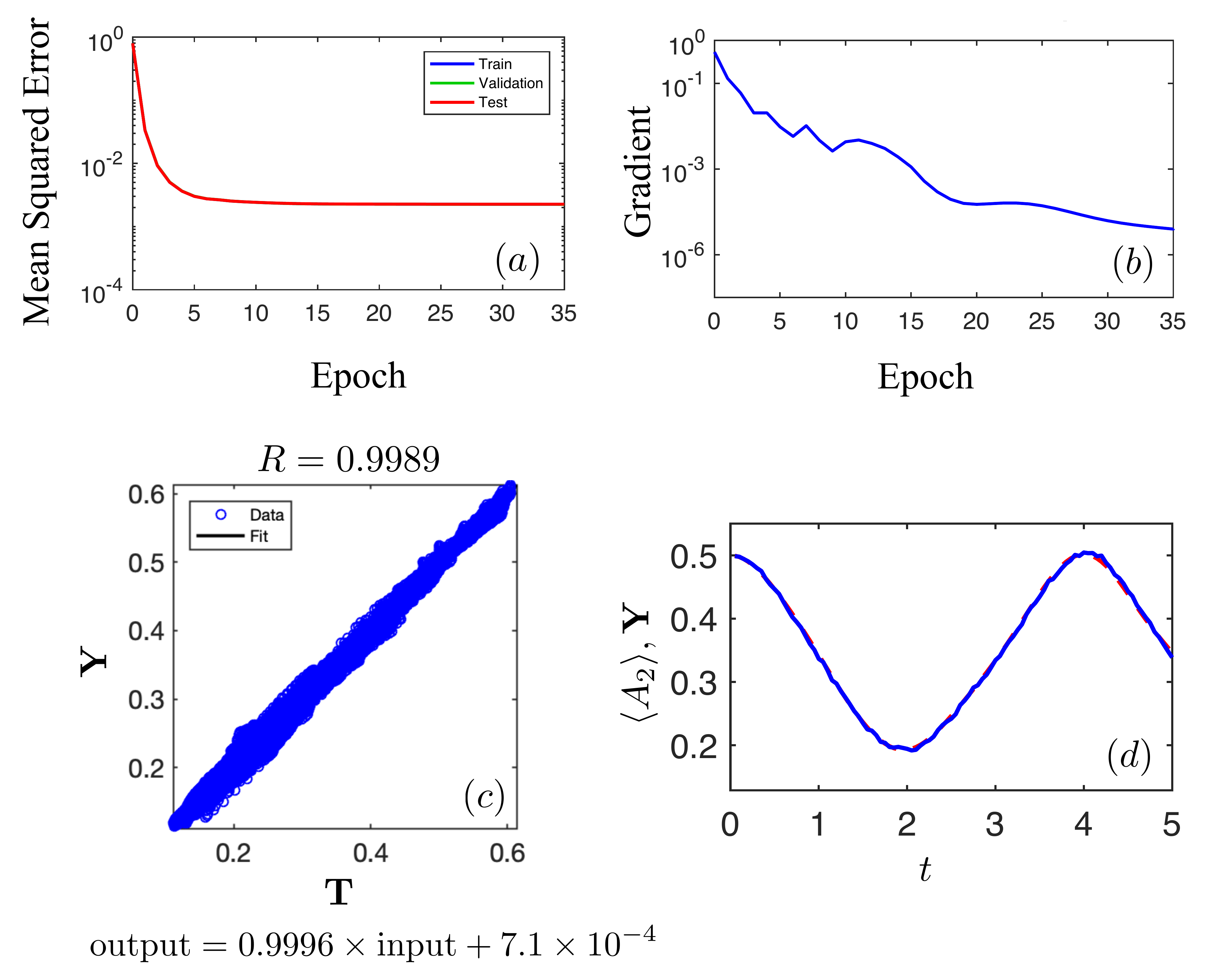}}
		\caption{\label{Fig3} (a-c)  Training results of the NN to establish the relation 
$F_2(g_x, g_y)_{j=1, ..., N_T} \approx \langle \hat{A}_2(t_{j=1, ..., N_T}; g_x, g_y)\rangle$ in the presence of 2D magnetic field in $x-y$ plane and $N=6$ with a finite number of measurements, $N_m=100$: (a) Cost function or mean squared error; (b) gradient of the training set; (c) Regression of the outputs of  such a NN with respect to the exact expectation values (targets) $\textbf{T} = \langle \hat{A}_2(t_{j=1, ..., N_T}; g_x, g_y)\rangle$. The fit line (solid black), expressed in the equations shown below, approximately matches the ideal relation $\textbf{Y} = \textbf{T}$, while $R$ is the correlation coefficient of the outputs and the targets. (d) Comparison of the outputs from the NNs trained by the data with shot-noise measurements $N_m=100$ (solid blue) with the exact expectation value $\langle \hat{A}_2 \rangle$ obtained from the Hamiltonian $\hat{H}_2$ (dashed red), when $g_x = 0.0998$, $g_y=0.1236$. }
	\end{center}
\end{figure}

We now turn our attention to a more realistic scenario, where the calibration of the NN is performed using shot noise measurements. In order to simulate an experimental acquisition, we numerically draw a set of $N_m$ binary measurements $x_m^j$ with $m=1,\ldots, N_m$ per time instant $t_j$, where $x_m^j\in \{0,1\}$. The  outcomes are drawn from the probability of measuring the observable $\hat{A}_1$ at time time $t_j$ and the corresponding external field $g_x$, i.e. $\langle \hat{A}_1(t_j;g_x)\rangle$. That is, $x_m^j\sim B(1,\langle \hat{A}_1(t_j;g_x)\rangle)$ where $B(1,p)$ denotes the Bernoulli distribution. We then use $\Theta=\{g_x\}$ as the input of the NN, while the corresponding set of averaged values $d_j=\sum_{m=1}^{N_m}x_m^j/N_m$ as the target dataset to train the NN, i.e. ${\bf T}=\{d_1,\ldots,d_{N_T}\}$. As in the previous case, we consider $N_T=101$ time instances in the interval $0\leq t\leq 5$, so that the number of input and output neurons of the NN are, as before, $1$ and $101$, respectively.  Training the NN with shot-noise measurements introduces statistical fluctuations. On the one hand, for a large number of measurements $N_m\ll 1$ the shot-noise averaged value converges to the exact expectation value $d_j\approx \langle \hat{A}_1(t_j;g_x)\rangle$. A large  number of measurements, however, requires long experimental runs and thus leads to a large overhead in experimental resources. On the other hand, a reduced number of measurements may speed up the calibration stage at the expense of introducing further errors in the approximation of the dynamics by the NN, and thus in the estimation precision. In the following we set a reasonable number of measurements per time instant, $N_m=100$, for which the training provides good results. We have tested cases with other $N_m$; in particular for $N_m=50$ one still finds a good accuracy of the NN, while for $N_m=20$ measurements, shot noise dominates and it deteriorates  the performance of the NN.   Moreover, in order to mitigate the impact of statistical fluctuations in the NN, we perform $10$ repetitions per calibrated value of $g_x$, so the whole dataset contains $n=51\times10$ examples from which $70\%$, $15\%$, $15\%$ represent the training, validation and test sets. The resulting regression of the NN considering a finite number of measurements is shown in Fig.~\ref{Fig2} (e). The NN trained with data including shot noise provides a high accuracy regression with respect to the ideal targets $\langle \hat{A}_1(t_{j=1, ..., N_T}; g_x) \rangle$. Taking a randomly chosen value $g_x = 0.1253$, the outputs from this NN (solid blue)  coincide well with the exact numerical result calculated from the Hamiltonian $\hat{H}_1$ (dotted black) (cf. Fig.~\ref{Fig2} (f)). The NN is therefore capable of establishing the relation given in Eq.~(\ref{F-A-gx}), so that the NN can be used to compute the likelihood, cf. Eq. (\ref{likelihood-NN}), and thus to perform Bayesian inference.

\subsection{2D external magnetic field}
We consider now an extension of the previous case, namely, a ferromagnetic XXZ spin-1/2 chain under a 2D magnetic field in $x-y$ plane we want to estimate. The Hamiltonian of the system can be written as 
\begin{eqnarray}
\label{H-2D}
\hat{H}_2(g_x, g_y) = - \sum_{i=1}^{N-1} (\hat\sigma_i^x \hat\sigma_{i+1}^x + \hat\sigma_i^y \hat\sigma_{i+1}^y+J_z\hat{\sigma}_{i}^z\hat{\sigma}_{i+1}^z) + g_x\sum_{i=1}^N \hat\sigma_i^x + g_y\sum_{i=1}^N \hat\sigma_i^y. 
\end{eqnarray}
Again, the initial state $\hat{\rho}_0$ is taken as the ground state for $g_x=g_y=0$, and we consider $J_z=0$. 
Letting the system evolve from the initial state for some time $0\leq t\leq 5$, we measure the local observable $\hat{A}_2 = (\hat{\sigma}^y_{N/2} +1) / 2$ for different values of $g_x$ and $g_y$ in the range $0\leq g_{x}\leq 0.5$ and $0\leq g_{y}\leq 0.5$. For the considered examples, the local magnetization along the $y$ axis provides a better sensitivity than along $x$ or $z$ axes.

For the calibration of the NN we proceed as before, including shot-noise measurements. Recall, however, that now the inputs $\Theta = \{g_x, g_y\}$ require $2$ neurons for the input layer of the NN. We train the NN so that the outputs ${\bf Y}$ reproduce the targets that are the averaged value shot-noise measurements of $\hat{A}_2$, $\textbf{T} = \{d_1,\ldots,d_{N_T}\}$. As before, we consider $N_T=101$, while the total number of datasets becomes $n=51^2\times 10$ due to the estimated parameters in two dimension and the $10$ repetitions for mitigating the impact of shot-noise fluctuations. In this case, the trained NNs allow us to find the relation
\begin{eqnarray}
\label{F-A-2D}
F_2(g_x, g_y)_{j=1,..., N_T} \approx \langle A_2(t_{j=1, ..., N_T}; g_x, g_y) \rangle.
\end{eqnarray}
See Fig.~\ref{Fig3} for the NN training results  to reproduce the dynamics of the system with $N=6$ spins for different values of $g_x$ and $g_y$.



\section{Estimation precision}\label{s:scaling}

The quantum Fisher information (QFI)~\cite{Braunstein94,Paris09} provides the ultimate bound to the achievable precision of a parameter $\theta$ encoded in a quantum state $|\Psi_\theta \rangle$ over all possible quantum measurements. The QFI is defined as 
\begin{eqnarray}
\label{QFI}
I_\theta^{\rm{QFI}}  = 4 \left( \left\langle \partial_\theta \Psi | \partial_\theta \Psi \right\rangle -
|\langle \Psi | \partial_\theta \Psi \rangle |^2 \right),
\end{eqnarray}
where $|\partial_\theta \Psi\rangle$ denotes the derivative of the state with respect to the parameter $\theta$.  The minimum variance  for an unbiased estimator of $\theta$ for a single measurement is set by the Cramer-Rao bound,
\begin{eqnarray}
\label{variance}
(\Delta \theta)^2 \geq (\Delta \theta^{\rm{QFI}})^2 = \frac{1}{I_\theta^{\rm QFI}}.
\end{eqnarray} 
The scaling with the number of probes or particles $N$ is determined by the QFI. Recall that standard quantum limit fixes the precision scaling as $\Delta \theta\propto N^{-1/2}$, while the Heisenberg limit refers to a better scaling of the form $\Delta \theta\propto N^{-1}$.

In Sec.~\ref{s:NN} we have shown how the NN allows us to faithfully reproduce the dynamics of the quantum many-body system upon a calibration stage using finite number measurements. Hence, we can compute the likelihood from a set of observations ${\bf D}$, and so the posterior distribution over the relevant parameters, $\Theta=\{g_x\}$ or $\Theta=\{g_x,g_y\}$ for the one or two-dimensional external field, respectively. The posterior distribution follows from the Bayes' theorem, $P(\Theta|{\bf D})\propto P({\bf D}|\Theta)P(\Theta)$, and we take the prior distribution as flat or uninformative in the considered range for $g_x$ and $g_y$ (cf. Sec.~\ref{s:NN}). The estimated parameter and its variance can be computed from its marginal posterior distribution, i.e. for $\theta_j$ we have as $\theta_j^{\rm{est}} = \int d\theta_j \theta_j P(\theta_j | \textbf{D})$ and $(\Delta \theta_j^{\rm{est}})^2 = \int d\theta_j (\theta_j - \theta_j^{\rm{est}})^2 P(\theta_j |\textbf{D})$, where the marginal distribution is given by $P(\theta_j | \textbf{D}) = \int \Pi_{i \neq j} d\theta_i P(\Theta | \textbf{D})$.

\begin{figure}[t]
	\begin{center}
		\scalebox{0.35}[0.35]{\includegraphics{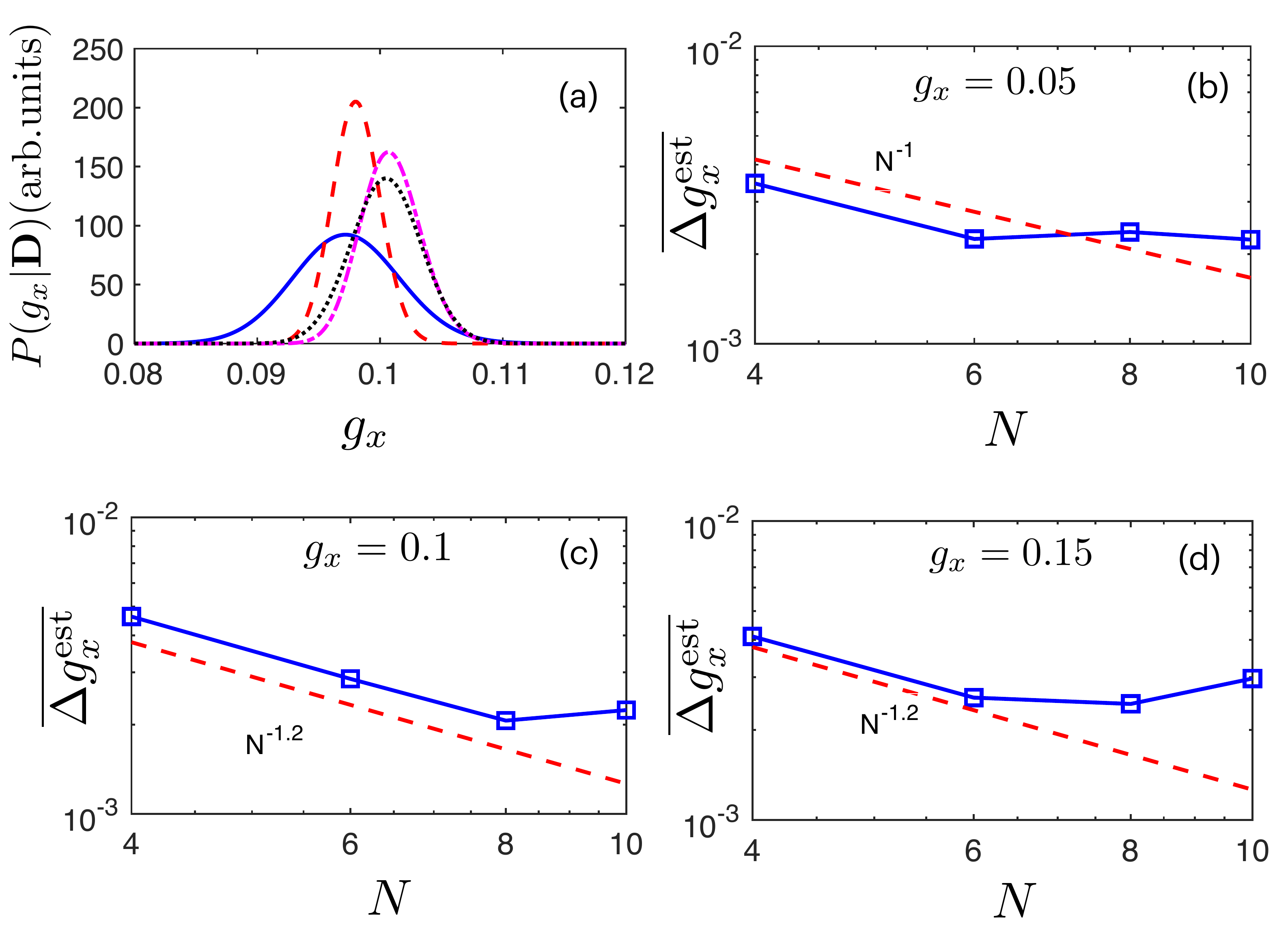}}
		\caption{\label{Fig4} (a) Posterior probability distributions of the one-dimensional external field strength $g_x$ for a $N=4$ (solid, blue), $N=6$ (dotted, black), $N=8$ (red, dashed) and $N=10$ (dot-dashed, magenta) computing  $P(\textbf{D} | g_x)$ assisted by the NN. The target value corresponds to $g_x=0.1$. (b-d) The scaling of estimation precision (squared, solid blue). The red dashed lines provide a guide for the eye with various scalings.}
	\end{center}
\end{figure}

\begin{figure}[t]
	\begin{center}
		\scalebox{0.35}[0.35]{\includegraphics{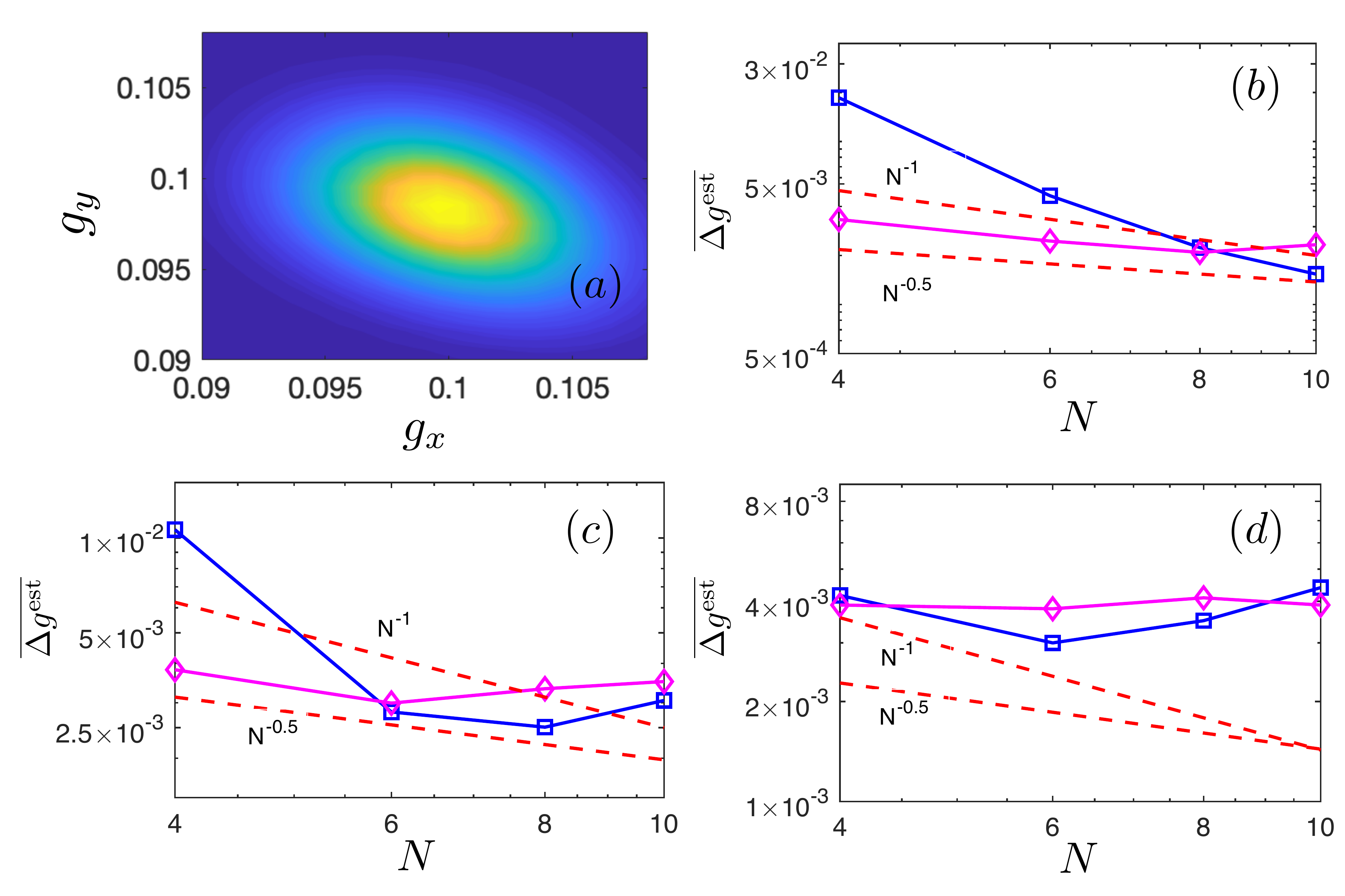}}
		\caption{\label{Fig5} (a) Posterior probability distributions for $N=6$ in terms of $g_x=0.1$ and $g_y=0.1$
		using the likelihood $P(\textbf{D} | g_x, g_y)$ assisted by the NN.  (b-d) The scaling of estimation precision of $\overline{\Delta g_x^{\rm{est}}}$ (squared, solid blue) and $\overline{\Delta g_y^{\rm{est}}}$ (diamond, solid magenta) for different $g_x$ and $g_y$, where the value with each $N$ is derived from Bayesian inference with the likelihood assisted by the NN. The red dashed lines indicate the standard quantum and Heisenberg limit scalings, i.e. $N^{-0.5}$ and $N^{-1}$.}
	\end{center}
\end{figure}

Employing the trained NNs, we compute the resulting mean and standard deviations for different target values $g_x$ (and $g_y$ for the two-dimensional case) simulating an experimental acquisition ${\bf D}$. This experimental acquisition is simulated in the same manner as when generating the datasets for the calibration step, cf. Sec.~\ref{s:NN}, albeit with an arbitrary number of measurements $N_p$ per time instant. Without lack of generality, we set $N_p=N_m=100$. Note that the precision scaling with the number of particles $N$ is not significantly altered by considering different number of measurements $N_p$.  This results in $g_x^{\rm est}$ and $\Delta g_x^{\rm est}$ (and $g_y^{\rm est}$ and $\Delta g_y^{\rm est}$). Moreover, in order to compute a statistically significant precision, we average the resulting standard deviation for $10$ different observations under the same target values. In this manner we find $\overline{\Delta g_x^{\rm est}}$ (and $\overline{\Delta g_y^{\rm est}}$ for the two-dimensional case). 

In Fig.~\ref{Fig4} we summarize the results to determine the strength of a one-dimensional external field $g_x$ using the NNs trained with the datasets with $N_m=100$ measurements. In Fig.~\ref{Fig4}(a) we show the posterior distribution for an increasing number of spins for $g_x=0.1$, while the other panels of Fig.~\ref{Fig4} show the scaling of the averaged standard deviation for three different values of $g_x$, i.e. $0.05$, $0.1$ and $0.15$. The specific values for $g_x^{\rm est}$ and $\Delta g_x^{\rm est}$, as well as their averaged values over $10$ repetitions can be found in \ref{app:tab}.  Remarkably, the behavior of the standard deviation for small sizes indicates a Heisenberg ($\Delta g_x\propto N^{-1}$) or even a super-Heisenberg scaling ($\Delta g_x\propto N^{-\alpha}$ with $\alpha>1$), although the scaling breaks down for some $N$. Note that the larger $g_x$, the smaller the size where the scaling breaks. These issues have been previously discussed~\cite{Rams18} for the same model. Indeed, as $g_x$ gets larger the system is pushed out of the critical region ($g_x=0$) where the sensitivity is higher, reducing the precision as more probes are included  to finally saturate (see~\cite{Rams18} for more details). In addition, it is worth noting that fixing a maximum time $t_f$ for the dynamical protocol ($0\leq t\leq t_f=5$) also limits the precision for magnetic-field strengths such that $t_f g_x\ll 1$.

For the estimation of the strength of a two-dimensional magnetic field we proceed in a similar manner. In Fig.~\ref{Fig5}(a) we show the bi-dimensional posterior distribution $P(g_x,g_y|{\bf D})$ for $N=6$ spins with $g_x=g_y=0.1$. In the other panels of Fig.~\ref{Fig5}, the behavior of the averaged standard deviation with an increasing number of spins is shown. As for the one-dimensional case, the model presents an improvement in the precision for small sizes and strengths, which is captured by our inference protocol. The numerical values can also be found in~\ref{app:tab}.

Besides these expected issues, it is important to stress that this hybrid metrological scheme incorporating the NN to perform Bayesian inference is able to capture the quantum-enhanced scaling in the precision as long as the quantum many-body system used as a sensor allows for it.

\section{Conclusion}\label{s:con}
In this article we propose a protocol to design a quantum many-body magnetometer using Bayesian inference techniques based on the ability of neural networks to faithfully reproduce the dynamics of the quantum many-body system. Bayesian inference techniques require the computation of the microscopic dynamics of the quantum many-body system under different parameters. For a general quantum many-body system, the computation of the dynamics soon becomes impractical even for small system sizes. To overcome this obstacle, we make use of a suitably trained neural network that is able to reproduce its dynamics. The neural network enables then an efficient Bayesian inference of external magnetic fields. In this manner, the neural network circumvents the need for a microscopic model description of the system at the expense of requiring a calibration step to establish a relation between dynamics and external magnetic fields. The calibration of the NNs could be performed solely based on experimental data, thus opening the door to employ experimental quantum many-body systems as sensors that are beyond the classical simulation capabilities. We illustrate this procedure in a XXZ model that allows for Heisenberg or even super-Heisenberg scaling in the precision in a parameter regime.
Our numerical results show that the external field strength can be obtained with high accuracy even for a finite number of measurements. This protocol showcases the advantage of Bayesian inference making use neural networks for parameter estimation, and thus paves the way to use quantum many-body systems as sensors under realistic conditions while benefiting from quantum resources to improve their precision.

\section*{Acknowledgments}
J.~C. acknowledges the Ram\'{o}n y Cajal   (RYC2018-025197-I) research fellowship, the financial support from Spanish Government via EUR2020-112117 and Nanoscale NMR and complex systems (PID2021-126694NB-C21) projects, the EU FET Open Grant Quromorphic (828826), and the Basque Government grant IT1470-22. This work was supported in part by the Basque Government through the ELKARTEK Program under Grant KK-2022/00062, “QFIRST - Dispositivos en Tecnolog\'i{a}s Cu\'{a}nticas, and under Grant KK-2022/00041, “BRTA QUANTUM: Hacia una especializaci\'{o}n armonizada en tecnolog\'{i}as cu\'{a}nticas en BRTA”.

\appendix 
\section{Neural network reproducing the dynamics including decoherence}\label{app:decoh}

\begin{figure}[t]
	\begin{center}
		\scalebox{0.3}[0.3]{\includegraphics{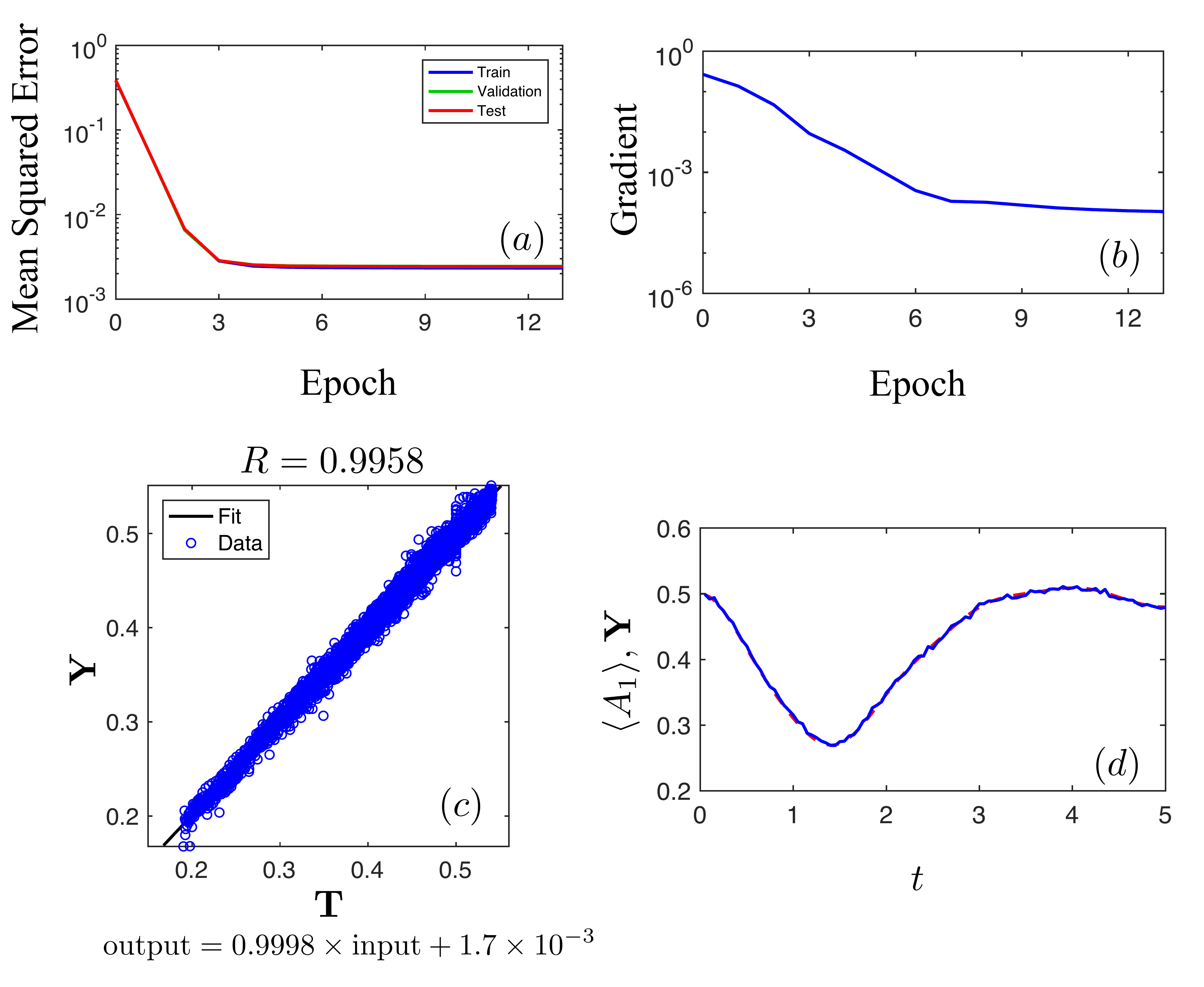}}
		\caption{\label{FigA1} (a-c)  Training results of the NN to establish the relation 
$F_1(g_x)_{j=1, ..., N_T} \approx \langle \hat{A}_1(t_{j=1, ..., N_T}; g_x)\rangle$ for 1D magnetic field and $N=4$ with a finite number of measurements $N_m=100$ and decoherence with rate $\gamma=1/10$: (a) mean squared error; (b) gradient of the training set; (c) regression of the outputs of  such a NN with respect to the ideal targets $\textbf{T} = \langle \hat{A}_1(t_{j=1, ..., N_T}; g_x)\rangle$. The fit line (solid black) demonstrates that $\textbf{Y} = \textbf{T}$ to a good approximation, while $R$ is the correlation coefficient of the outputs and the targets. (d) Comparison of the outputs from the NNs trained by the data with $N_m=100$ (solid blue) in accordance with the exact numerical calculation $\langle \hat{A}_1(t;g_x)\rangle$ (dashed red) from the Lindblad master equation, Eq.~(\ref{DM}) when $g_x= 0.2053$. }
	\end{center}
\end{figure}

As an example to stress the generality of the neural network to reproduce the dynamics of the quantum many-body system with imperfections, we illustrate the performance when including decoherence. In this case, the considered that dynamics of the system is described by means of a Lindblad master equation~\cite{Lindblad76,Breuer}
\begin{eqnarray}
\label{DM}
\frac{d\hat{\rho}}{dt} = -i [\hat{H}, \hat{\rho}] +\sum_k \gamma_k \left(\hat{L}_k \hat{\rho} \hat{L}_k^{\dagger} - \frac{1}{2}\left\{\hat{L}_k^{\dagger} \hat{L}_k, \hat{\rho} \right\} \right).
\end{eqnarray}
For simplicity, we consider equal decoherence rates $\gamma = \gamma_k$ for all qubits and the spin operator $\hat{L}_k = \hat{\sigma}_{-}$. Due to the large computational cost to simulate Eq.~(\ref{DM}), we access system with $N=4$ and $6$ spins for the 1D magnetic field, setting a moderate decoherence rate $\gamma=1/10$. In both cases the trained NN reproduces the exact dynamics to a very good approximation. 
The training process of such a NN is equivalent to one described in Sec.~\ref{s:NN}. In Fig. \ref{FigA1}, we show the training results of performance, gradients and regression of the outputs of the NN with respect to the exact numerical results including decoherence.  Inputting one randomly chosen example $g_x=0.2053$, which does not belong to any dataset, we find the outputs from the NN that corresponds to the targets $\langle \hat{A}_1 (t_{j=1, ..., N_T}; g_x) \rangle$ to a very good approximation. Note that the precision in estimating $g_x$ may be spoiled depending on the decoherence rate $\gamma$, but the NN still allows for an efficient Bayesian estimation. This hybrid scheme can achieve quantum-enhanced scaling in the precision as long as the quantum many-body system used as a sensor allows for it.

\section{Numerical results}\label{app:tab}
In the following Tab.~\ref{gx-1D-1times},~\ref{gx-1D} and~\ref{gx-2D} we provide the numerical values that are plotted in Figs.~\ref{Fig4} and~\ref{Fig5} for the estimated magnetic-field strengths $g_x$, $g_y$, their corresponding standard deviations as well as their averaged values over $10$ repetitions.

\begin{table}[t]
	\centering
	\begin{tabular}{lccc}
	        
		\hline
		Target $g_x$& $N$  & $g_x^{\rm{est}}$ & $\Delta g_x^{\rm{est}}$  \\
		\hline
		0.1 & 4 & 0.09729 & 0.004343 \\ 
		& 6 & 0.09749 & 0.002931 \\ 
		& 8 & 0.09817 & 0.001949 \\ 
		& 10 & 0.09824 & 0.002105 \\
		\hline
	\end{tabular}
\caption{\label{gx-1D-1times}
Estimated magnetic field strength $g_x$ and its standard deviation $\Delta g_x^{\rm{est}}$ for $N=4, 6, 8, 10$ spins obtained from Bayesian inference employing the trained NN with a finite-number of measurements. The simulated observations ${\bf D}$ have been obtained for a target value $g_x=0.1$.}
\end{table}

\begin{table}[]
	\centering
	\begin{tabular}{lccc}
		\hline
		Target $g_x$& $N$  & $\overline{g_x^{\rm{est}}}$ & $\overline{\Delta g_x^{\rm{est}}}$ \\ 
		\hline
		0.05 & 4 & 0.04788 & 0.004513 \\ 
		& 6 & 0.04872 & 0.002251 \\ 
		& 8 & 0.04769 & 0.002376 \\ 
		& 10 & 0.04931 & 0.002239 \\ 
		\hline
		0.1 & 4 & 0.1005 & 0.004513 \\ 
		& 6 & 0.1005 & 0.002850 \\ 
		& 8 & 0.1006 & 0.002062 \\ 
		& 10 & 0.1003 & 0.002238 \\ 
		\hline
		0.15 & 4 & 0.1529 & 0.004112 \\ 
		& 6 & 0.1511 & 0.002566 \\ 
		& 8 & 0.1515 & 0.002444 \\ 
		& 10 & 0.1498 & 0.002973 \\ 
		\hline
		0.2 & 4 & 0.2016 & 0.003225 \\ 
		& 6 & 0.2001 & 0.002625 \\ 
		& 8 & 0.2013 & 0.003672 \\ 
		& 10 & 0.2015 & 0.004255 \\ 
		\hline
	\end{tabular}
\caption{\label{gx-1D}
Averaged values for the estimated magnetic-field strength $\overline{g_x^{\rm{est}}}$ and for its standard deviation $\overline{\Delta g_x^{\rm{est}}}$ for $10$ repetitions and for different system sizes.}
\end{table}

\begin{table}[]
	\centering
	\begin{tabular}{lccccc}
		\hline
		Target $g_x$, $g_y$ & $N$  & $\overline{g_x^{\rm{est}}}$ & $\overline{g_y^{\rm{est}}}$ & $\overline{\Delta g_x^{\rm{est}}}$ & $\overline{\Delta g_y^{\rm{est}}}$ \\ 
		\hline
		0.05, 0.05 & 4 & 0.04153 & 0.04909   & 0.01859  &   0.003323 \\ 
	         	& 6 & 0.04752 & 0.04999 &   0.004651 &  0.002452 \\  
	         	& 8 & 0.04798 & 0.04881 & 0.002235 &  0.002088 \\ 
	         	& 10 & 0.04923 & 0.04983 &0.001535 & 0.002327 \\ 
		\hline
		0.075, 0.075 & 4 & 0.06909 & 0.07339 & 0.01062 & 0.003817 \\ 
	         	& 6 & 0.07590 & 0.07662 & 0.002804 &  0.002991 \\ 
	         	& 8 & 0.07505 & 0.07477 & 0.002509 &  0.003322 \\ 
	         	& 10 & 0.07343 & 0.07687 &  0.003050 & 0.003502 \\ 
		\hline 
		0.1, 0.1 & 4 & 0.09779 & 0.09799 & 0.004168 & 0.003942 \\ 
		      & 6 & 0.1002 & 0.09752 & 0.003011 & 0.003913 \\ 
	              & 8 & 0.1023 & 0.09630 & 0.003542 & 0.004176 \\ 
		     & 10 & 0.1022 & 0.09813 &  0.004438 & 0.003917 \\ 
		\hline
		\end{tabular}
\caption{\label{gx-2D}
Averaged values for the estimated magnetic-field strengths $\overline{g_x^{\rm{est}}}$, $\overline{g_y^{\rm{est}}}$ and for their standard deviations $\overline{\Delta g_x^{\rm{est}}}$, $\overline{\Delta g_y^{\rm{est}}}$.}
\end{table}

\clearpage

\section*{References}

\providecommand{\newblock}{}

\end{document}